# Improved vector and scalar masses in the massive Schwinger model

C. Adam*

Wolfgang-Pauli Institute, c/o Institute of Mathematics, Vienna University, 1090 Vienna

## Abstract

The lowest ("vector") and next-lowest ("scalar") bound-state masses of the massive Schwinger model have been determined recently to a very high accuracy numerically on the lattice. Therefore, improved results for these bound-state masses from analytical calculations are of some interest. Here, we provide such improved results by employing both standard and renormal-ordered (fermion) mass perturbation theory, as well as a consistency condition between the two perturbative calculations. The resulting bound-state masses are in excellent agreement with the lattice results for small and intermediate fermion mass, and remain within 10% of the exact results even in the limit of very large fermion mass.

*)email address: adam@mat.univie.ac.at

# 1 Introduction

The massive Schwinger model is QED in 1+1 dimensions with one massive fermion species [1]–[5]. It is, one the one hand, the simplest non-trivial gauge field theory. On the other hand, it nevertheless shares some properties with more complicated gauge field theories like, e.g., QCD in 3+1 dimensions. The model has, for instance, a non-trivial vacuum [1, 2], and the fermions are confined in the sense that they do not show up as physical particles in the particle spectrum [2, 6].

The Lagrangian density of the model is

$$L = \bar\Psi(i\slashed{\partial} - g\slashed{A} - m)\Psi - \frac{1}{4}F_{\mu\nu}F^{\mu\nu}. \qquad (1)$$

Here, both the fermion mass $m$ and the coupling constant $g$ have the dimension of a mass, therefore the dimensionless parameter $m/g$ effectively controls the behaviour of the theory.

One interesting feature of the model is the formation of bound states. For small coupling (i.e., large $m/g$), the bound states are just fermion–antifermion bound states that are weakly bound by the 1+1 dimensional Couloumb potential [2]. Therefore, the masses of the bound states are $2m$ plus some small (positive) binding energies and should approach $2m$ in the limit of vanishing coupling $g$.

For large coupling, the model is best described in an equivalent bosonic formulation [1, 2]. In fact, for a massless fermion ($m = 0$), the model may be solved exactly, and it is equivalent to the theory of one free, massive boson field ("Schwinger boson") with Schwinger mass $\mu_0 = g/\sqrt\pi$ [7, 8]. For small but non-zero fermion mass ($m \neq 0$), this Schwinger boson turns into an interacting particle, and its mass acquires corrections which can be calculated perturbatively in $m$ [9, 10, 11]. This Schwinger boson may be identified with the lowest bound state, the vector state. Further, a bound state of two Schwinger bosons is formed within the bosonic formulation of the theory [2, 11, 12], and this bound state of two bosons may be identified with the next lowest bound state, the scalar state.

These two lowest bound states remain stable for all values of the parameter $m/g$. In general, all higher bound states become unstable for sufficiently large coupling (small $m/g$) [2]. For vanishing vacuum angle ($\theta = 0$) there exists a third stable bound state [2], which may be interpreted as a bound state of one Schwinger boson and one two-boson bound state within a bosonic formulation [11]. Here we will consider the case of vanishing vacuum angle (because all lattice data are for this case), but we will be concerned only with the two lowest bound states, because to our knowledge there are no lattice data available for the third bound state.

For the calculation we will use perturbative expansions of the bosonic formulation of the theory for different choices of the normal-ordering mass, together with some consistency conditions among these different perturbation expansions. Qualitatively this has already been discussed in [13], therefore we will describe the calculation only briefly before presenting explicit results.

## 2 Vector mass calculation

The theory of Eq. (1) is equivalent to the bosonic theory [1, 2, 5] (for conventions we refer to [13])

$$L = -N_{\mu_0}\left[\frac{1}{2}\Phi(\Box - \mu_0^2)\Phi + \frac{e^\gamma}{2\pi}\mu_0 m\left(\cos(\sqrt{4\pi}\Phi) - 1\right)\right]. \tag{2}$$

Here $\mu_0 = g/\pi^{1/2}$ is the Schwinger mass of the massless ($m = 0$) Schwinger model, $\gamma = 0.5772$ is the Euler constant and $N_\mu$ denotes normal-ordering w.r.t. $\mu$. Further, a constant vacuum energy density has been subtracted, as usual.

By making use of the well-known normal-ordering relation

$$N_{\bar\mu}\, e^{\pm i\beta\Phi(x)} = \left(\frac{\mu}{\bar\mu}\right)^{\frac{\beta^2}{4\pi}} N_\mu\, e^{\pm i\beta\Phi(x)} \tag{3}$$

(see [14]), we may rewrite the Lagrangian density (2) for arbitrary normal-ordering mass $\bar\mu$ like

$$L = -N_{\bar\mu}\left[\frac{1}{2}\Phi(\Box - \mu^2)\Phi + \frac{e^\gamma}{2\pi}\bar\mu\, m\left(\cos(\sqrt{4\pi}\Phi) - 1\right)\right]. \tag{4}$$

By shifting the quadratic (in $\Phi$) part from the interaction part $\sim \cos\sqrt{4\pi}\Phi$ to the free part of the Lagrangian (4), and by insisting that the normal-ordering mass is equal to the total mass of the free Lagrangian, we arrive at

$$L = -N_{\bar\mu_0}\left[\frac{1}{2}\Phi(\Box - \bar\mu_0^2)\Phi + \frac{e^\gamma}{2\pi}m\bar\mu_0\left(\cos(\sqrt{4\pi}\Phi) + 2\pi\Phi^2 - 1\right)\right] \tag{5}$$

with

$$\bar\mu_0 = e^\gamma m + \sqrt{\mu_0^2 + e^{2\gamma}m^2}. \tag{6}$$

The resulting Lagrangian (5) has a slightly more complicated interaction term (containing a $\Phi^2$ term in addition to the $\cos\sqrt{4\pi}\Phi$ term), and the dimensionless expansion parameter is now

$$\lambda \equiv \frac{m}{\bar\mu_0}. \tag{7}$$

This new expansion parameter $\lambda$ has the advantage that it tends to a finite and rather small value even for large $m/g$.

The lowest bound-state mass $M_1$ (the "vector" state mass or Schwinger mass) may be calculated from the two-point function $\langle\Phi(x)\Phi(y)\rangle$ for both Lagrangians (2) and (5) by a perturbation expansion in the respective interaction terms. Via a Fourier transformation the mass pole may be determined. For ordinary mass perturbation theory (i.e., from the Lagrangian (2)), the result up to second order is

$$(M_1^{m,2})^2 = \mu_0^2 + 2e^\gamma\,\mu_0 m + e^{2\gamma}m^2(A_+ + A_-), \tag{8}$$

where

$$A_\pm = \int_0^\infty dr\, r\left[e^{-2K_0(r)} - 1 + I_0(r)(\pm e^{\mp 2K_0(r)} \mp 1 + 2K_0(r))\right] \tag{9}$$

$$A_+ = -0.6599, \qquad A_- = 1.7277 \tag{10}$$

see e.g. [9, 10, 11]. The notation $M_1^{m,2}$ means that the subscript 1 stands for the first (lowest) bound state mass, and the superscript $m, 2$ stands for the $m$ perturbation expansion up to second order. Here, $\mu_0^2$ is the order zero result, and there are both first and second order contributions. This result describes the vector state mass quite well for small fermion mass $m$, see Fig. 1, but it diverges, of course, in the limit of very large $m$.

Starting from the Lagrangian (5), the order zero result is $\bar\mu_0$. There is no first order contribution (because there is no $\Phi^2$ term in the interaction Lagrangian), and the total mass up to second order is

$$(M_1^{\lambda,2})^2 = \bar\mu_0^2 + e^{2\gamma} m^2 (A_+ + A_-) - 2e^{2\gamma} m^2$$
$$= \mu^2 + 2e^\gamma \cos\theta\, m\sqrt{\mu^2 + e^{2\gamma} \cos^2\theta\, m^2} + e^{2\gamma} m^2 (A_+ \cos 2\theta + A_-), \tag{11}$$

see [13]. Please observe that one piece of the second order (in $\lambda$) contribution, namely the term $-2e^{2\gamma} m^2$, has precisely cancelled the $o(m^2)$ piece of the lowest order (in $\lambda$) contribution, $\bar\mu_0^2$. This cancellation ensures that $(M_1^{\lambda,2})^2$, when Taylor expanded in $m$ up to second order, coincides with the second order result of mass perturbation theory, $(M_1^{m,2})^2$. $(M_1^{\lambda,2}/m)$ reaches a finite value (which is about 55% above the true value 2) in the limit of large $m$. Further, $M_1^{\lambda,2}$ describes the data quite well for small $m/g$, but already for $m/g = 1$ it is about 30% above the true value. Therefore, a further improvement would be desireable. This aim can be achieved by requiring that the result of the $\lambda$ perturbation expansion, when Taylor-expanded in $m$, should agree with the result from the $m$ perturbation expansion also in higher orders. This requirement is perfectly sensible, because $\lambda$ has a well-defined Taylor expansion into $m$ to all orders, and the physical results from both formulations of the theory should agree after all. A further Taylor expansion in $m$ of the mass $(M_1^{\lambda,2})^2$ contains, e.g., the term

$$\bar\mu_0^2|_{m^3} = e^{3\gamma} \frac{m^3}{\mu_0}, \tag{12}$$

in third order in $m$, and infinitely many higher order terms. Now the crucial point is that such terms cannot be produced by the perturbation theory in $m$, because Wick contractions of the interaction term $\cos\sqrt{4\pi}\Phi$ all must contain integrals of a special type of transcendental functions (the $E_\pm$ functions of [11, 13]). Therefore, the higher orders in the $\lambda$ perturbation expansion must contain terms that precisely cancel terms like (12) when Taylor-expanded in $m$, as we have already observed for the second order contribution. The term (12) is, e.g., precisely cancelled by the third order term (in $\lambda$ perturbation expansion)

$$\delta(M_1^{\lambda,3})^2 = -e^{3\gamma} \bar\mu_0^2 \lambda^3 = -e^{3\gamma} \frac{m^3}{\bar\mu} \tag{13}$$

which must, therefore, be present in the $\lambda$ perturbation expansion. This cancellation has to continue in higher orders, and the Taylor coefficients of higher order terms, like (13), have to be cancelled, too. This cancellation condition enables us to perform a further partial higher order calculation in $\lambda$ perturbation expansion up to some order $N$ by employing the equation

$$(M_1^{\lambda,p})^2 \equiv (M_1^{\lambda,2})^2 - \bar{\mu}_0^2 \sum_{l=3}^{N} \frac{e^{l\gamma}}{l!} c_l \lambda^l \stackrel{!}{=} (M_1^{m,2})^2 + O(m^{N+1}) \tag{14}$$

where a Taylor expansion in $m$ up to order $N$ is understood on the l.h.s. The coefficients $c_l$ can be determined recursively, which can be easily done with the help of Mathematica. We show the result for $N = 30$ in Fig. 1 and in Table 1. As can be seen, the improvement for intermediate values of $m/g$ is dramatic. For $m/g = 1$, e.g., the result deviates from the lattice result by less than 0.5%. In the limit of very large $m$, the deviation from the exact result $2m$ is about 8%.

Within $\lambda$ perturbation theory, the higher order terms that we have found from condition (14) correspond to disconnected products of Wick contractions, where in each connected component one $\cos\Phi$ term is contracted with one or several $\Phi^2$ terms of the interaction Lagrangian. There exist other types of terms which do not appear in the $m$ perturbation expansion (contractions of two or more $\cos\Phi$ with at least one $\Phi^2$), but these may be detected in a way similar to the above argument only when compared with higher orders in $m$ perturbation theory. Unfortunately, already a full third order calculation (both in $m$ and in $\lambda$ perturbation theory) is rather involved.

## 3   Scalar mass calculation

The scalar mass may be calculated from the four-point function $\langle \prod_{l=1}^{4} \Phi(x_l) \rangle$ by focusing on the $s$-channel contribution and by again determining the mass pole. Here, we want to use the results of the last section for the vector mass, i.e., we want to use the masses $(M_1^{m,2})^2$ and $(M_1^{\lambda,p})^2$ for the respective perturbation theories, because they are better approximations to the physical masses than the bare masses $\mu_0$ resp. $\bar{\mu}_0$. However, naively using these masses within the Lagrangians (2) and (5) would lead to the problem that the propagating masses are no longer equal to the normal-ordering masses, which would produce tadpole contributions. Therefore, we change the normal-ordering masses in both Lagrangians and use for the subsequent calculations the Lagrangians

$$L = -N_{M_1}\left[\frac{1}{2}\Phi(\Box - M_1^2)\Phi + \frac{e^\gamma}{2\pi}M_1 m \left(\cos(\sqrt{4\pi}\Phi) - 1\right)\right]. \tag{15}$$

and

$$L = -N_{M_1}\left[\frac{1}{2}\Phi(\Box - M_1^2)\Phi + \frac{e^\gamma}{2\pi}mM_1\left(\cos(\sqrt{4\pi}\Phi) + 2\pi\Phi^2 - 1\right) + e^\gamma m(\bar{\mu}_0 - M_1)\Phi^2\right] \tag{16}$$

where $M_1$ will be finally chosen equal to $M_1^{m,2}$ in the first case and equal to $M_1^{\lambda,p}$ in the second case. Please observe that the $\Phi^2$ part of the interaction Lagrangian in (16) is *not* changed by the renormal-ordering and that we do not introduce a further shift such that all $\Phi^2$ contributions belong to the free part of the Lagrangian. Here we just remark that this choice is perfectly legitimate, i.e., the resulting perturbation theory is as well-defined as the others. Later on it will become clear why this choice is even advisable for our purposes.

Now, we have to calculate the bosonic four-point function, perform the Fourier transformation, focus on the $s$-channel contribution and find the mass pole in the kinematic variable $s$. The calculation is essentially equivalent to the calculations within mass perturbation theory that have been done in [12, 11], therefore we just quote the result. One finally arrives at the pole mass equation (or gap equation) for the scalar mass $M_2$ in leading order

$$1 = \frac{e^\gamma m}{M_1} \frac{4}{y^2} \frac{1}{\sqrt{\frac{4}{y^2}-1}} \arctan \frac{1}{\sqrt{\frac{4}{y^2}-1}} \tag{17}$$

where $y = M_2/M_1$. It turns out that in leading order the gap equation is formally the same for both perturbation theories, the only difference being the different choices for $M_1$. The $M_1$ masses have been determined fully up to second order in the respective perturbation expansions. It turns out that the above gap equation induces a leading order contribution to the mass ratio $y$ which is of second order (see Eq. (19) below), therefore it is consistent to use the gap equation in leading order only. Further, we understand now why we had to choose the Lagrangian (16) as it stands. When $M_1 = M_1^{m,2}$ in (further renormal-ordered) $m$ perturbation theory and $M_1 = M_1^{\lambda,p}$ in (further renormal-ordered) $\lambda$ perturbation theory, then it is automatically guaranteed that the resulting masses $M_2$ will agree up to order $N$ when Taylor-expanded in $m$. Here we should be somewhat clearer about the Taylor expansion. The point is that we want to use the gap equation (17) directly for a numerical evaluation, which is then no longer accessible to a Taylor expansion. However, it is easy to generate an expansion in the coupling constant from the gap equation (17). Just rewrite it as

$$\epsilon = 4 \frac{e^\gamma m}{M_1} \frac{1}{\sqrt{4-\epsilon^2}} \operatorname{arccot} \frac{\epsilon}{\sqrt{4-\epsilon^2}}, \quad \epsilon^2 \equiv 4 - y^2 \tag{18}$$

and solve this equation iteratively, starting with $\epsilon = 0$ on the right-hand side. For the resulting expressions, we demand that they must agree up to order $N$ when Taylor-expanded in $m$.

There are several reasons for using the gap equation (17) directly for a numerical evaluation instead of a further expansion (iteration of Eq. (18)). Firstly, the leading second order contribution in $m$ has a rather large coefficient [12, 11, 15],

$$\frac{M_2}{M_1} \simeq 2 + 24.625 \frac{m^2}{g^2} \tag{19}$$

and, therefore, approximates the lattice results rather badly already for quite small values of $m$. Secondly, after a further expansion, it is not even guaranteed that $y < 2$ (as it has to be for a bound state), whereas $y < 2$ holds automatically for the gap equation (17).

Remark: Please observe that for an iteration of Eq. (18) in leading order, as has been done in [12, 11], the further normal-ordering of Eq. (15) is unnecessary, because it only changes results in higher orders.

The gap equation has been evaluated numerically with the help of Mathematica. Already within mass perturbation theory (i.e., for $M_1 = M_1^{m,2}$) the results of the numerical solution are quite reasonable up to $m/g \sim 1/2$ and are much better than the leading order result (19). For the Lagrangian (16), i.e., for $M_1 = M_1^{\lambda,p}$, the results are shown in Table 1, again for the choice $N = 30$. The results agree very well with the lattice data for small $m$. For $m/g \sim 1$, the discrepancy is about 4%, and in the limit of very large $m$ the deviation from the exact result $M_2 = 2m$ is about 9%.

## 4  Summary

In this paper we have performed an improved calculation of the two lowest bound-state masses of the massive Schwinger model (the vector and the scalar bound-state mass). For this purpose we used the bosonized version of the theory with two different normal-ordering prescriptions and with two slightly different interaction terms in the Lagrangian, see Eqs. (2) and (5). In addition, we used a consistency condition between the two perturbation theories in order to partially determine higher order contributins to the second ($\lambda$) perturbation theory, see Eq. (14). The resulting masses agree very well with the results from recent lattice calculations [15, 16] for small and intermediate values of the parameter $m/g$. In the limit of very large $m/g$ they deviate from the known exact results by less than 10%. There are additional results on the masses $M_1$ and $M_2$ in the literature, but we chose to compare with the lattice results of [15, 16] as they seem to be the most accurate numerical estimates which are available up to now. For additional results on these masses with various methods, please consult [15] and the literature cited there (some further references can also be found in [13]).

In principle, the results obtained in this paper could be further improved by methods that are analogous to those used in this paper. However, any further improvement requires a full third order calculation of the vector mass $M_1$ within mass perturbation theory, which is quite difficult to do, as already mentioned. Further, the methods emlployed in this paper may, in principle, be used for the calculation of further observables in the massive Schwinger model, like, e.g., higher masses or decay widths and scattering cross sections. Also, it should be possible to use a variant of this renormal-ordered perturbation theory for a discussion of the multi-flavour Schwinger model where ordinary mass perturbation theory fails due to infrared divergencies [17, 18].

**Acknowledgment:** The author acknowledges support from the Austrian START award project FWF-Y-137-TEC of N.J. Mauser. Further thanks are due to M. Frank for help with the Mathematica programming.

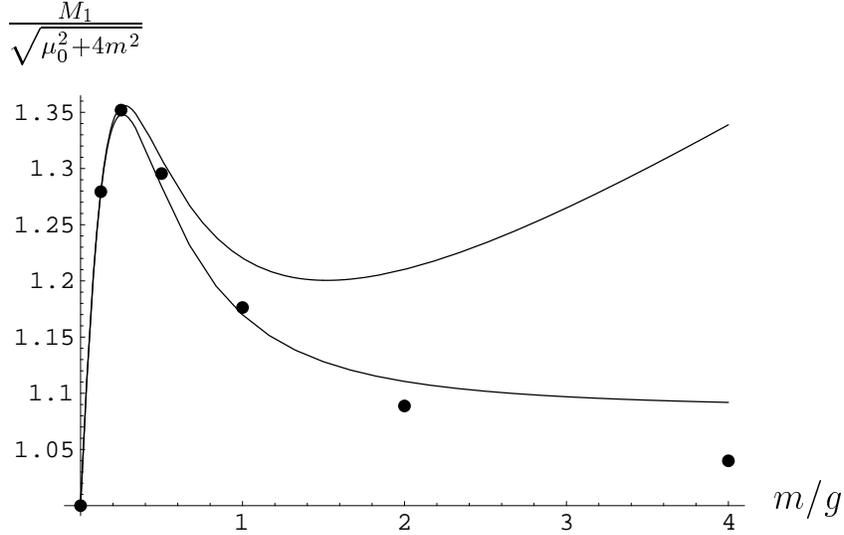

Figure 1: The vector mass $M_1$, normalized such that the exact value is 1 in both limits $m \to 0$, $m \to \infty$. The dots are the lattice results of [16], the upper curve is the result $M_1^{m,2}$ of mass perturbation theory, the lower curve is the improved result $M_1^{\lambda,p}$ with $N = 30$.

Table 1: Comparison of vector and scalar masses $M_1/g$, $M_2/g$ as functions of $m/g$. Our results are compared with the finite-lattice estimates of [15, 16]. For $m = 0$, "this work" are the known exact results.

| $m/g$ | $M_1/g$ this work | $M_1/g$ lattice, [16] | $M_2/g$ this work | $M_2/g$ lattice, [15] |
|---|---|---|---|---|
| 0 | 0.564 | 0.564 | 1.128 | 1.11 |
| 0.125 | 0.789 | 0.789 | 1.470 | 1.47 |
| 0.25 | 1.016 | 1.019 | 1.730 | 1.74 |
| 0.5 | 1.474 | 1.487 | 2.165 | 2.20 |
| 1 | 2.431 | 2.444 | 2.988 | 3.12 |
| 2 | 4.486 | 4.398 | 4.876 | 5.00 |
| 4 | 8.757 | 8.340 | 9.072 | 8.85 |
| 8 | 17.40 | 16.29 | 17.78 | 16.68 |
| 16 | 34.75 | 32.24 | 35.41 | 32.56 |
| 32 | 69.46 | 64.19 | 70.71 | 64.45 |